\documentstyle[12pt,epsfig]{article}
\newcommand{\be}{\begin{equation}}\newcommand{\ee}{\end{equation}}
\newcommand{\bea}{\begin{eqnarray}}\newcommand{\eea}{\end{eqnarray}}
\newcommand{\nn}{\nonumber}
\newcommand{\Real}{\Re e}
\newcommand{\Imm}{\Im m}
\newcommand{\wt}{\widetilde}
\newcommand{\gsim}{\stackrel{>}{_\sim}}

\begin{document}

\thispagestyle{empty}
\renewcommand{\thefootnote}{\fnsymbol{footnote}}
\begin{center}
{\hfill LNF-96/053(P)}\vspace{0.2cm} \\
{\hfill hep-ph/9610328}\vspace{2cm} \\
\vglue 1.0 true cm
{\Large The role of chiral loops in $\eta\to\pi^0\pi^0\gamma\gamma$}
\footnote{Work supported in part by HCM, EEC-Contract No. 
CHRX--CT920026 (EURODA$\Phi$NE).}
\vspace{1.5cm} \\
\renewcommand{\thefootnote}{\alph{footnote}}
\setcounter{footnote}{0}
{\bf S. Bellucci}\footnote{E-mail: bellucci@lnf.infn.it} {\bf and 
G. Isidori}\footnote{E-mail: isidori@lnf.infn.it}
                    \vspace{1cm} \\
INFN-Laboratori Nazionali di Frascati, P.O. Box 13, 00044 Frascati,
            Italy
\vspace{2.5cm} \\
{\bf Abstract} 
\end{center}
\narrower{We consider the rare 
decay $\eta\to\pi^0\pi^0\gamma\gamma$ and calculate
the non--resonant contribution to the amplitude to one loop in
Chiral Perturbation Theory. We display our result as both a diphoton energy
spectrum and a partial decay rate as a function of the photon energy cut.
It turns out that the one--loop correction can be
numerically very important and could be detected, at sufficiently large
center--of--mass photon energies, from a measurement of 
the partial decay width.}
\vfill
\begin{center}
October 1996
\end{center}
\setcounter{page}0
\renewcommand{\thefootnote}{\arabic{footnote}}
\setcounter{footnote}0
\newpage

\section{Introduction}
                      
The $\eta\to\pi\pi\gamma\gamma$ decays been been recently 
analyzed \cite{a1} in the framework of Chiral Perturbation
Theory (CHPT) \cite{Wein79,GL} (see \cite{Ecker} for a recent review). 
 The experimental
interest for such rare decays stems from the large number of observed
$\eta$'s anticipated at various $\eta$--factories, e.g. CELSIUS, ITEP
and DA$\Phi$NE \cite{dafne}, 
as well as at other facilities, such as GRAAL, MAMI,
ELSA and CEBAF. 

In the neutral decay $\eta\to\pi^0\pi^0\gamma\gamma$ 
we can identify two physically distinct contributions to the 
amplitude: 
\be
A(\eta\to\pi^0\pi^0\gamma\gamma) = A_R + A_{NR}~.
\ee
The first contribution (that we call the `resonant amplitude')
is characterized by the $\pi^0$ pole in the  
diphoton invariant mass squared ($s_{\gamma\gamma}$) and is 
proportional to the on--shell 
amplitudes $\eta\to3\pi^0$ and 
$\pi^0\to\gamma\gamma$:\footnote{~The minus sign in Eq.~(\ref{prima})
is due to our convention for the amplitudes.} 
\be
A_R =  - \frac{A( \eta\to 3 \pi^0 ) A(\pi^0 \to \gamma\gamma )  }{ 
             s_{\gamma\gamma}-m_{\pi^0}^2 }~.    \label{prima}
\ee  
By construction, $A_R$ can be predicted up to a phase
from the experimental data on $\eta\to3\pi^0$ and  
$\pi^0\to\gamma\gamma$ and needs not to be calculated in CHPT. 
The main result of \cite{a1}, in the $\eta\to\pi^0\pi^0\gamma\gamma$ 
channel, can be summarized as follows: the resonant contribution dominates 
over the tree--level non--resonant one in the whole phase space.  
                            
The non--resonant amplitude cannot be predicted using experimental data
and must be calculated in CHPT. In Ref.~\cite{a1} $A_{NR}$ 
has been calculated at the tree level, i.e. to $O(p^4)$;
at this order only the $\eta$--exchange diagram ($\eta\to \pi^0\pi^0\eta^* \to 
\pi^0\pi^0\gamma\gamma$) contributes. Kn\"ochlein et al. have 
considered  also the $\eta'$--exchange diagram 
(formally of higher order in CHPT),  but for both $\eta$-- and 
 $\eta'$--exchange they found 
 that $A_{NR}$ is negligible with respect to $A_R$. 
This is because the lowest--order 
$\eta\eta\pi^0\pi^0$ and $\eta\eta'\pi^0\pi^0$
vertices vanish in the limit $m_u=m_d=0$. 
In spite of the analogous suppression factor in the
$\pi^0$--exchange contribution, which is proportional to $m_u-m_d$,
the enhancement due to the pole
makes $A_R$ dominant with respect to $A_{NR}$ 
over the full range of kinematical parameters. 
However, there are good reasons to presume that this suppression 
does not occur to one loop. Indeed, recently Talavera et al. 
\cite{a2} have shown that the one--loop contribution to the
$\gamma\gamma\to\pi^0\pi^0\pi^0$ amplitude -- which is obviously connected
to the $\eta\to\pi^0\pi^0\gamma\gamma$ one  -- dominates over
the tree--level result (the former is one order of magnitude larger
than the latter). The reason for this enhancement can be  
traced to the fact that the tree--level amplitude is proportional to 
$m_{\pi}^2$, whereas the one--loop
correction is not affected by such suppression.

In this Letter we compute $A_{NR}$ to one loop, 
 neglecting isospin--breaking effects and the suppressed
 $\eta$--exchange diagrams. In this limit only 
one--particle--irreducible (1PI) diagrams contribute and the 
result is finite. Furthermore, $A_{NR}$ does not receive any contribution
from the $O(p^6)$ counterterms analyzed in \cite{a4}, just as it
happens in the $\gamma\gamma\to\pi^0\pi^0\pi^0$ case. We find that
the one--loop contribution to $A_{NR}$ is dominant, with respect to the
corresponding tree--level one.
Once again, the reason is that the tree--level amplitude
goes to zero in the limit $m_u=m_d=0$, whereas the one--loop result
does not vanish in this limit. We find, in addition, that there exists
a kinematical region in the phase space, i.e. a region of sufficiently
large values of $s_{\gamma\gamma}$, where the resonant amplitude
-- which is a background that shadows $A_{NR}$ --
is suppressed, and a measurement of the decay width
$\Gamma (\eta\to\pi^0\pi^0\gamma\gamma )$ would allow us to detect 
a purely $O(p^6)$ effect.

The outline of this Letter is as follows.
We begin in Section 2 with the description of the kinematical
variables for the decay $\eta\to\pi^0\pi^0\gamma\gamma$ and a list
of interaction terms relevant to the calculation of the one--loop
amplitude. 
In Section 3, after recalling the expression of the decay amplitude
at the tree level, we describe the calculation of the one--loop corrections
to the non--resonant amplitude and give the analytic expression of the
result. 
We proceed in Section 4 to calculate the decay width, starting from
the sum of $A_R+A_{NR}$. We display the result in the form of both a diphoton
energy spectrum and a partial decay rate as a function of the energy cut
around $s_{\gamma\gamma}^{1/2}=m_{\pi^0}$.
Then we determine the
phase--space region, in terms of a $s_{\gamma\gamma}$  range, where
the suppression of the background due to $A_R$ may allow to detect the
one--loop effects in $A_{NR}$.
We end the Letter with some concluding remarks and a short
discussion of further developments, including possible extensions
of the one--loop calculation to the charged pions channel, as well as
to the reactions $\gamma\gamma\to\eta\pi\pi$.

\section{Kinematical variables and interaction terms}

The kinematics of the decay 
$\eta(q) \to \pi^0(p_1) \pi^0(p_2) \gamma(k_1,\epsilon_1)
\gamma(k_2,\epsilon_2)$ can  be described in terms of five independent 
scalar variables which we choose as:
\bea
&s_{\pi\pi}=(p_1+p_2)^2~, \qquad &z_{1,2}=k_{1,2}\cdot(p_1+p_2)~, \label{p12} \\
&s_{\gamma\gamma}=(k_1+k_2)^2~, \qquad  &z_3=(k_1+k_2)\cdot(p_1-p_2)~.
\eea
If we write the decay amplitude in the following way
\be
A(\eta\to\pi^0\pi^0\gamma\gamma) = 
e^2 \epsilon_1^\mu \epsilon_2^\nu A_{\mu\nu}~, 
\ee
the decay width is given by
\be
\Gamma(\eta\to\pi^0\pi^0\gamma\gamma) = \frac{\alpha^2_{\rm em} }{2^{11}
\pi^6 m_\eta } \int \frac{d^3p_1}{p_1^0} \frac{d^3p_2}{p_2^0}
\frac{d^3k_1}{k_1^0} \frac{d^3k_2}{k_2^0}
\delta^{(4)}(p_1+p_2 + k_1+k_2 ) A^{\mu\nu}A^*_{\mu\nu}~.\label{GammaT}
\ee
\medskip

Since the process $\eta\to\pi^0\pi^0\gamma\gamma$ 
involves the electromagnetic interaction of an odd
number of pions, the decay amplitude receives contributions 
only from the odd--intrinsic parity sector of CHPT
and thus is at least $O(p^4)$. 

The CHPT lagrangian, expanded up to $O(p^4)$, is given by
\be
{\cal L}={\cal L}^{(2)}+{\cal L}^{(4)}~,
\ee
where 
\be
{\cal L}^{(2)} = {F^2 \over 4 } \mbox{tr}\left( D_\mu U D^\mu U^\dagger
+ \chi U^\dagger  + \chi^\dagger U \right) \label{due}
\ee
and ${\cal L}^{(4)}$ can be split into the odd--intrinsic
anomalous part (i.e. the Wess--Zumino term \cite{a5}) and 
the $O(p^4)$ Gasser--Leutwyler lagrangian \cite{GL}
\be
{\cal L}^{(4)}={\cal L}_{WZ}+\sum_{i=1}^{10}L_i{\cal L}^{(4)}_{i}~.
\label {L4}
\ee
As usual, we assume the exponential parametrization 
$U=\mbox{exp}(i\sqrt{2}P_8/F)$, where  $P_8$
is the $SU(3)$ 
octet matrix of pseudoscalar mesons and $F$ coincides to
the lowest order with the 
charged pion decay constant $F_{\pi}=92.4$~MeV \cite{GL,a6}.
The covariant derivative in Eq.~(\ref{due}) is given by 
$D_\mu U = \partial_\mu U +i e A_\mu [Q,U] $, where 
$A_\mu$ is the photon field and $Q=$diag$(2/3,-1/3,-1/3)$.
Finally, we employ the identification 
$\chi =\chi^{\dagger}=2 B{\cal M}$ in the external scalar sources, where
 ${\cal M}=\mbox{diag}(m_u,m_d,m_s)$ is the quark mass matrix and 
 $B$ can be identified to the lowest order with the mass 
ratio $B_0=m_{\pi}^2 / (m_u+m_d)$.

In principle, in this decay one could take into account also 
the $\eta-\eta^{'}$ mixing, i.e. the mixing of 
$P_8$ with the singlet--field $\eta_0$. However, as shown in 
\cite{a1}, this effect can be safely neglected in the non--resonant
amplitude. Hence in the loop calculation
we identify the mass--eigenstate $\eta$ with the octet field
$\eta_8$. In the same spirit, since we neglect isospin--breaking 
effects in $A_{NR}$, in the following we assume $m_\pi=m_{\pi^0}$.       

\begin{figure}[t]
\centerline{\epsfig{file=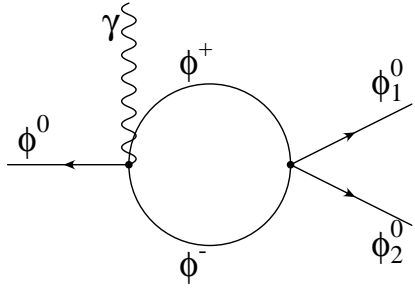,height=4cm}}
\caption{1PI one--loop diagrams for the 
$\phi^0 \to \phi_1^0 \phi_2^0 \gamma\gamma$ transition. 
The second photon line  has to be 
attached to the charged lines running in the loop and to the vertices.} 
 \protect\label{fig1}
\end{figure}

For the interaction terms necessary to calculate the tree--level amplitudes 
one can refer to \cite{a1}. However, not just for the sake of completeness,
but also given that we use a small subset of the couplings in \cite{a1},
we collect them in the following:
\begin{eqnarray}
A^{(2)}(\eta_8\to \pi^0\pi^0\pi^0) & = &  3 A^{(2)}(\eta_8\to \pi^0\pi^+\pi^-)
= \frac{B_0(m_u-m_d)}{\sqrt{3}F_{\pi}^2}~,  \\
A^{(2)}(\eta_8\to \eta_8 \pi^0\pi^0) & = &  A^{(2)}(\eta_8 \to \eta_8
\pi^+\pi^-) = \frac{B_0(m_u+m_d)}{3 F_{\pi}^2}~, \label{eepp2}\\
A^{(4)}(\pi^0\to\gamma\gamma ) & = & \sqrt{3} A^{(4)}(\eta_8\to\gamma\gamma ) =
 \frac{e^2}{4\pi^2F_{\pi}} \epsilon_{\mu\nu\alpha\beta}
\epsilon_1^{\mu}k_1^{\nu}\epsilon_2^{\alpha} k_2^{\beta}~.
\end{eqnarray}
In addition, in order to compute the one-loop diagrams in fig.~\ref{fig1},
we introduce the generic couplings
\begin{eqnarray}    
A^{(2)}(\phi^+\phi^-\to\phi_1^0\phi_2^0 ) & = &  a s_{\pi\pi} +
bm_{\pi}^2 + c (p_+^2 -m_{\pi}^2)+d(p_-^2-m_{\pi}^2)~, \\ 
A^{(4)}(\phi^0\to\phi^+\phi^-\gamma ) & = & 
f \epsilon_{\mu\nu\alpha\beta}\epsilon^{\mu}k^{\nu}p_+^{\alpha}q^{\beta}~,
\label{couplings}
\end{eqnarray}
where $q$, $p_{\pm}$, $p_{1,2}$ and $k$ are the (outgoing) momenta of  
the pseudoscalars $\phi^0$, $\phi^{\pm}$, $\phi_{1,2}^0$ and of the photon,
 respectively. The constants
$a,b,c$ and $d$ have the dimensions of inverse mass squared,
whereas $f$ has those of an inverse 
mass cubed. As we will show in the next section,
the `off--shell couplings' $c$ and $d$ are irrelevant for our
process, since their contribution to the amplitude cancels out
as a consequence of the gauge invariance. In the 
$\pi^+\pi^-\to \pi^0\pi^0$ and $\eta_8\to \pi^+\pi^-\gamma$ cases, useful
in order to estimate the dominant pion loops, we find
\be
a=-b=\frac{1}{F_{\pi}^2}\qquad\mbox{and} \qquad 
f = -\frac{e}{4\sqrt{3}\pi^2 F_{\pi}^3}~.
\ee

\setcounter{equation}0
\section{The decay amplitude to one loop}

In this Section we would like first to briefly recall the 
expression of the tree--level amplitudes $A_R^{(4)}$ and
$A_{NR}^{(4)}$, obtained by considering the
$\pi^0$ and $\eta_8$ exchange diagrams \cite{a1}
\bea
A^{(4)}_R=-\frac{e^2}{4\sqrt{3}\pi^2 F_{\pi}^3}
\frac{B_0(m_u-m_d)}{(s_{\gamma\gamma}-m_{\pi^0}^2)}
\epsilon_{\mu\nu\alpha\beta}\epsilon_1^{\mu}k_1^{\nu}\epsilon_2^{\alpha}
k_2^{\beta}~,\label{tree} \\
A^{(4)}_{NR}=-\frac{e^2}{12\sqrt{3}\pi^2 F_{\pi}^3}
\frac{B_0(m_u+m_d)}{(s_{\gamma\gamma}-m_{\eta}^2)}
\epsilon_{\mu\nu\alpha\beta}\epsilon_1^{\mu}k_1^{\nu}\epsilon_2^{\alpha}
k_2^{\beta}~.\label{tree2} 
\eea
As already stated in the introduction, 
the enhancement factor due to the pion pole 
makes $A^{(4)}_R$ dominant  --in spite of its suppression factor
$(m_u-m_d)$-- with respect to $A_{NR}^{(4)}$ in the
entire kinematical space.

The $O(p^6)$ loop and counterterm (CT) contributions can be divided in 
three gauge--invariant 
subgroups: reducible $\pi^0$--exchange diagrams, reducible 
$\eta_8$--exchange diagrams and 1PI diagrams. 
\begin{enumerate}

\item[i.] The $\pi^0$--exchange diagrams, which include both loops and CT, 
contribute mainly to $A_{R}$. In principle such diagrams generate also 
a contribution to $A_{NR}$. Indeed, if we decompose the $\eta\to \pi^0
\pi^0(\pi^0)^*$ 
amplitude as follows:
\be
A(\eta\to \pi^0 \pi^0(\pi^0)^*)=A_{on-shell}(\eta\to 3\pi^0)
+  (s_{\gamma\gamma}-m_{\pi}^2)\times A_{off-shell}~,
\ee
the ${\cal A}_{off-shell}$ term drops out of $A_R$.
However this non--resonant contribution vanishes in the limit $m_u=m_d$ 
and thus can be safely neglected. On the other hand, 
$|{\cal A}_{on-shell}|$ can be extracted by means of
experimental data and  we do not need to evaluate it in CHPT.

\item[ii.] Also the $\eta_8$--exchange diagrams include both loops and CT.
These diagrams contribute only to $A_{NR}$ and can be 
safely neglected. Indeed, we have  explicitly  checked that 
the contribution of these diagrams is of the same order as the
tree--level result (\ref{tree2}) that is known to be small \cite{a1}.
The reason of such suppression  can be easily understood: the 
$\pi$-$\pi$ loops, which are expected to provide the dominant  
contribution, are suppressed by the factor $(m_u+m_d)$ in (\ref{eepp2})
just as the tree level result. There are contributions 
from the $K$-$K$ loops and ${\cal L}^{(4)}$  
which are not suppressed 
by $(m_u+m_d)$. They are nonetheless negligible, since we
are far below the kaon threshold and the CT combinations 
involved, i.e.~  $(L_1+L_3/6)$, $(L_2+L_3/3)$ and $L_4$,  
are small \cite{Ecker}. 

\item[iii.] The 1PI diagrams are the loop diagrams in fig.~1 (note that the
figure indicates schematically at least four distinct diagrams). The sum of
these contributions is finite and, as we will show in the following,
turns out to be the dominant contribution to $A_{NR}$.   

\end{enumerate}

The calculation of the loop diagrams in fig.~1 resembles a 
recent calculation by D'Ambrosio et al. \cite{a7} of
the radiative four--meson  amplitudes. The difference is that in our case 
one pseudoscalar field is replaced by one photon, but the main 
features of the result --simply dictated by QED-- are the same. 
We find indeed
\bea
A_{NR}^{1PI} &=& 4 e f  (as_{\pi\pi}+bm_\pi^2) \times \nn\\
&&\!\!\!\! \times \Biggr\{ \widetilde{ C_{20}}( s_{\pi\pi},-z_2) 
\epsilon_{\mu\nu\alpha\beta}
\epsilon_1^{\mu}k_1^{\nu} \left[(\epsilon_2 \cdot p_{12}) k_2^{\alpha}
 - z_2 \epsilon_2^\alpha
\right] q^{\beta}\; +\; (\epsilon_1,k_1 \leftrightarrow \epsilon_2,
k_2) \Biggr\}~, \label{1-l}
\eea
where $p_{12}=p_1+p_2$ and the 
function $\wt{C_{20}}(x,y)$ is given in appendix. 
As in \cite{a7}, gauge invariance forces the result to depend
 only from the `on--shell couplings' $a$, $b$ and $f$. 
Furthermore, the amplitude (\ref{1-l})
is $O(k_1,k_2)$ in the limit of vanishing photon momenta,
in analogy to the direct--emission amplitudes of \cite{a7} 
which are $O(k)$. 

Since we have written the two 
vertices in a general form, not only the dominant pion loops,
but also the kaon loops are represented in Eq.~(\ref{1-l}). 
From our general one--loop amplitude we recover, as a particular
case, part of the result of Talavera et al. \cite{a2} (i.e. the
contribution of 1PI diagrams). The precise correspondence between
the function   $\widetilde{ C_{20}}(x,y)$  and the
function $R(x,y)$ entering Eq.~(10) of \cite{a2} 
is given by
\bea
R(x,y)=32\pi^2 y \wt{ C_{20}}(x,y)~. \label{RC}
\eea

The amplitude $(\ref{1-l})$ depends only on the function 
$\widetilde{ C_{20}}$ and thus is finite.
This is a consequence of both the gauge invariance of the amplitude
and the fact that the on--shell $\pi^+\pi^-\to\pi^0\pi^0$  
amplitude depends only on $s_{\pi\pi}$ (i.e. it does not depend on the 
loop variables). We expect that the sum of 1PI diagrams is no more finite
 if the two external $\pi^0$'s are replaced by a $\pi^+$-$\pi^-$ pair.
Indeed, in this case not only the on--shell 
$\pi^+\pi^-\to\pi^+\pi^-$ amplitude does depend on the loop momenta,
but this sum is also not gauge invariant (in order to  obtain a 
gauge  invariant result, it is necessary to add the corresponding 
reducible diagrams with a photon emission from the external legs).    

\setcounter{equation}0
\section{Numerical analysis}

\begin{figure}[t]  
   \begin{center}
   \setlength{\unitlength}{1truecm}
       \begin{picture}(10.0,10.0)
       \epsfxsize 10.0 true  cm
       \epsfysize 10.0 true cm
       \epsffile{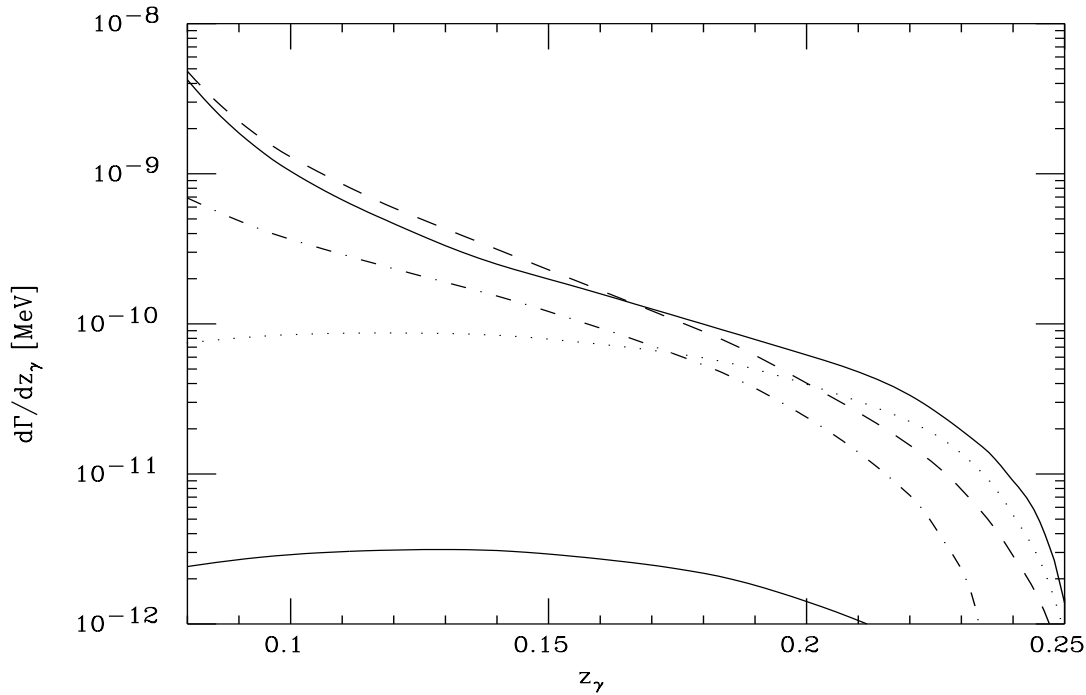}
   \end{picture}
   \end{center}
   \caption{Diphoton spectrum ($z_\gamma=s_{\gamma\gamma}/m_\eta^2$)
for the decay $\eta\to\pi^0\pi^0\gamma\gamma$. The upper full line is
the total contribution. The dashed line is the resonant contribution   
($|A^{phys}_R|^2$), the dotted line is the one--loop non-resonant 
contribution ($|A^{1PI}_{NR}|^2$) and the 
dash--dotted line is the asbolute value of
their interference ($\rho=2,~\alpha_0=0.18$). The 
lower full line is the tree-level non--resonant contribution 
($|A^{(4)}_{NR}|^2$).} 
   \protect\label{fig2}
\end{figure}
 
\noindent
The results of our analysis are summarized in figs. 2 and 3.
The plots have been obtained integrating numerically 
Eq.~(\ref{GammaT}) with the following decay amplitude:
\be
A(\eta\to\pi^0\pi^0\gamma\gamma) = A_R^{phys} +\left[ A^{(4)}_{NR}
+A^{1PI}_{NR} \right]~.  \label{sum}
\ee
The last two terms in Eq.~(\ref{sum}) are the CHPT results 
given in Eqs.~(\ref{tree2}) and  (\ref{1-l}),  whereas 
$A_R^{phys}$ denotes a phenomenological expression for
the resonant amplitude:
\be
A^{phys}_R=  A^{(4)}_R  \rho e^{i\alpha_0}~.  \label{Aphys} \\
\ee
The factor $\rho e^{i\alpha_0}$ in the above equation 
takes into account the corrections to the tree--level amplitude   
of $\eta\to 3\pi^0$, which are known to be large \cite{GLeta}. 
Assuming a flat Dalitz Plot for this decay --not withstanding experimental 
constraints-- and using the relation \cite{GLeta}
\be
B_0(m_d-m_u)= m^2_{K^0} - m^2_{K^+} - m^2_{\pi^0} + m^2_{\pi^+}~,
\ee
from the experimental data on $\Gamma (\eta\to 3\pi^0)$ \cite{a6} we find
$\rho = 2.0\pm 0.1$.

Contrary to $\rho$, the 
phase $\alpha_0$ cannot be extracted from the $\eta\to 3\pi^0$ data.
Similarly to the $K\to 3\pi$ analysis of \cite{DIPP}, in order 
to evaluate $\alpha_0$, we expand the one--loop 
$\eta\to 3\pi^0$ amplitude \cite{GLeta}  
around the center of the Dalitz Plot. Hence we obtain 
\be
\alpha_0 = \frac{ 1}{32 \pi F_{\pi}^2} \left( 1 - \frac{4 m_\pi^2 }{s_0} 
\right)^{1/2} (2 s_0 + m_\pi^2) \simeq 0.18~,
\ee
where $s_0=(m_\eta^2 +3m_\pi^2)/3$.

\begin{figure}[t]  
   \begin{center}
   \setlength{\unitlength}{1truecm}
       \begin{picture}(10.0,10.0)
       \epsfxsize 10.0 true  cm
       \epsfysize 10.0 true cm
       \epsffile{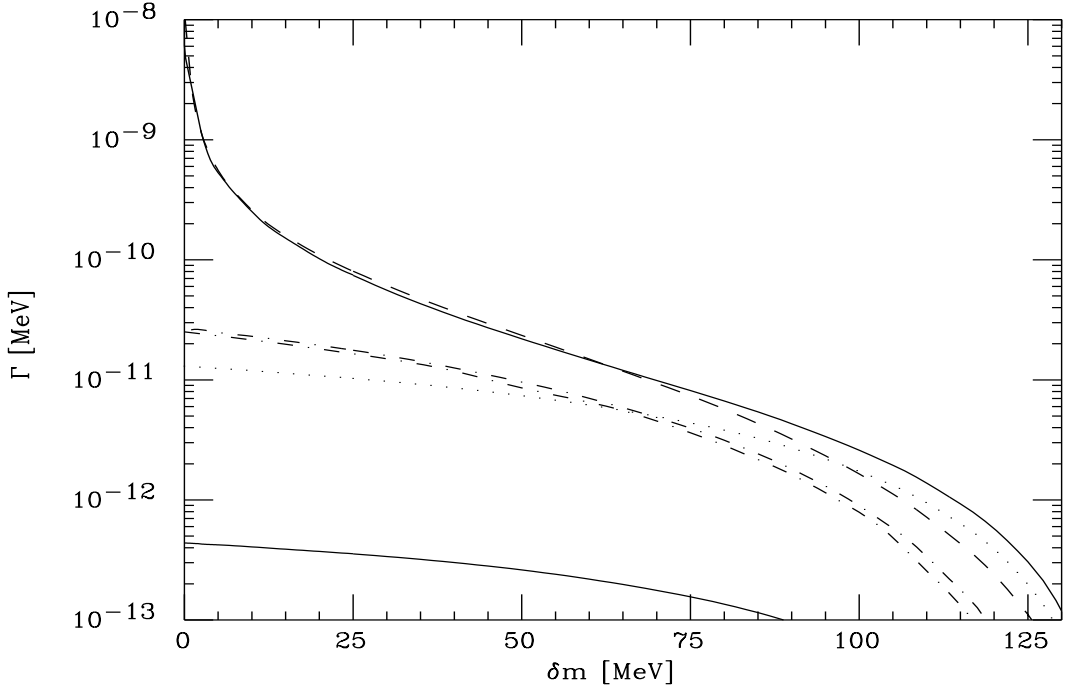}
   \end{picture}
   \end{center}
   \caption{Partial decay rate of 
$\eta\to\pi^0\pi^0\gamma\gamma$ 
as a function of the energy cut  
$|s_{\gamma\gamma}^{1/2}-m_{\pi^0}|<\delta m$. 
Full, dashed and dotted curves as in fig.~2. The two dash--dotted lines,
denoting the absolute value of the
interference between $A^{phys}_R$ and $A^{1PI}_{NR}$, 
have been obtained for $\alpha_0=0.16$ (upper line) and 
$\alpha_0=0.20$ (lower line).}
   \protect\label{fig3}
\end{figure}

As anticipated, the two figures show clearly that in the 
non--resonant amplitude the one--loop
result  dominates over the tree--level one in the 
whole phase space. Moreover, for $s_{\gamma\gamma} \gsim 0.15 m^2_\eta$
the non--resonant amplitude becomes non--negligible with respect to 
the resonant one. 
For $s_{\gamma\gamma} \gsim 0.20 m^2_\eta$
the dominant contribution to the decay is provided by $A_{NR}$.

All the distributions have been obtained using  
$\rho=2$ in $A_R^{phys}$ and 
considering the dominant $\pi$-$\pi$ loops only. 
We have explicitly checked by means of Eq.~(\ref{1-l})
that the kaon loops give a very small contribution, confirming the statement
of the previous Section that the $\pi$--$\pi$ loops dominate the
non--resonant amplitude in the whole phase space.
The dependence of our result on $\alpha_0$ is quite 
small, as shown in fig.~3, while a precise determination of $\rho$
is important, in order to determine with a good accuracy the region
where the non--resonant amplitude dominates over the resonant one.
Our analysis could be improved if
high precision data on  $\eta\to 3\pi^0$ were available.
They would allow, in principle, to include the (small)
$D$--wave contribution we neglected. However, even with the present
uncertainty on the normalization factor $\rho$, we can state that the loop
effect is in principle observable. Indeed, for
$s_{\gamma\gamma}> 0.23 m_{\eta}^2$, we
find that the non--resonant contribution is larger than the resonant one
by more than a factor of 2.

Notice that our distributions 
differ from those in \cite{a1}
 --apart from the one--loop corrections 
 to $A_{NR}$--  by an overall normalization 
factor. Since we agree with the analytic expressions reported in
\cite{a1}, this discrepancy can be traced to a problem in the program 
used to produce their plots. 

\setcounter{equation}0
\section{Discussion and outlook}

In this paper we have explicitly calculated the dominant one--loop 
corrections in CHPT to the decay $\eta\to\pi^0\pi^0\gamma\gamma$, 
going beyond the lowest order treatment of 
\cite{a1} -- the latter essentially corresponds to a simple
current algebra calculation. The phenomenological interest
of this process is due to the experimental facilities which can effectively
act, in the next few years, as $\eta$-factories. A similar physical motivation
led recently many authors to calculate leading corrections
to the lowest--order CHPT prediction in both decays,
such as $\eta\to\pi^0\gamma\gamma$ \cite{a8},
and scattering processes, such as
$\gamma\gamma\to\pi^+\pi^-$ \cite{a10} and
$\gamma\gamma\to\pi^0\pi^0$ \cite{a11}.
Earlier work on the CHPT predictions of pion polarizabilities can
be found in \cite{a14}, and the comparison 
with forward--angle dispersion sum--rules
is discussed in \cite{a15}.  Recently the charged--pion 
polarizabilities have been computed to two loops \cite{Burgi}.

Recent results on $\gamma\gamma\to\pi^0\pi^0\pi^0$
\cite{a2} have inspired us, since there the lowest--order
amplitude is suppressed and the corrections due to chiral loops
dominate the cross-section.
We found a similar result to hold for the non--resonant contribution
to the decay $\eta\to\pi^0\pi^0\gamma\gamma$.

Despite the enhancement due to the one--loop corrections, the
non-resonant amplitude is shadowed from the resonant (i.e. the 
$\pi^0$--exchange) contribution, over
a large portion of the diphoton spectrum. We have shown,
however, that for large
$s_{\gamma\gamma}$ the one--loop corrections
to the non-resonant amplitude dominate also over the resonant 
contribution (see figs.~2 and 3).
A measurement of the partial width of 
$\eta\to\pi^0\pi^0\gamma\gamma$ in this kinematical region --within the reach
of the future facilities-- would represent a new interesting test of
CHPT at $O(p^6)$.

We wish to conclude this Letter with a few comments concerning the
possibility of future developments. In particular, one might ask why we
concentrated on the neutral pion
channel, rather than calculating the amplitude of
$\eta\to\pi^+\pi^-\gamma\gamma$, which is statistically favored.
The reason is that the decay $\eta\to\pi^+\pi^-\gamma\gamma$ is dominated
by the bremsstrahlung of  $\eta\to\pi^+\pi^-\gamma$ \cite{a1}. 
Since the latter is not suppressed already at the tree level, 
we expect that the one--loop 
corrections not related to the  $\eta\to\pi^+\pi^-\gamma$ 
amplitude will be hardly detectable.\footnote{~From the theoretical point 
of view, in order to isolate these effects it is necessary to implement 
an appropriate definition of `generalized bremsstrahlung',
similarly to what has been done in \cite{a7} for the radiative--four--meson
amplitudes.}
From this point of view, when considering the outlook for a future calculation
and a potentially related measurement, we expect that the scatterings
$\gamma\gamma\to\pi^+\pi^-\eta$
and $\gamma\gamma\to\pi^0\pi^0\eta$ will provide more interesting
tools for the study of chiral--loop effects.

\section*{Added note}

After submitting this Letter, we learned about a similar calculation 
which included also the $\eta$--$\eta^{\prime}$ mixing effect and an 
estimate of $O(p^8)$ contributions in the resonant amplitude
\cite{bra}. This result confirms that the effect we 
calculated is the dominant one in the non--resonant amplitude. 

\section*{Acknowledgments}
S.B. wishes to thank E. Hourani and M. Knecht for the invitation
to the GRAAL Collaboration meeting at Orsay, where this
investigation was undertaken.
We also acknowledge useful discussions with Ll. Ametller, G. Colangelo,  
G. D'Ambrosio, G. Ecker and H. Neufeld. Special thanks are due to A. Bramon
for pointing out to us a sign error in an earlier version of this paper 
and informing us of his related work.

\setcounter{equation}0
\section*{Appendix}

The general definition of  
$\wt{C_{20}}(x,y)$ in terms of the 
three--denominator one--loop scalar functions 
can be found in \cite{a7}. 
In the $\pi$-$\pi$ case and for $x,\ x-2y > 4 m_\pi^2$ the 
explicit expression is given by:
\bea
(4\pi)^2 \Real \wt{C_{20}}(x,y) &=& {x \over 8 y^2}\left\lbrace 
\left(1-2{y\over x}\right)\left[\beta\log\left({1+\beta
\over 1-\beta}\right) -\beta_0\log\left({1+\beta_0
\over 1-\beta_0}\right)\right] \right. \nn\\ 
&& \left. \qquad + 
{m_\pi^2\over x} \left[\log^2\left({1+\beta_0 \over 1-\beta_0}
\right)-\log^2\left({1+\beta \over 1-\beta }\right) \right]
+ 2{y\over x} \right\rbrace~, \\
& & \nn \\
 (16\pi) \Imm \wt{C_{20}}(x,y) &=& - {x \over 8 y^2}\left\lbrace 
\left(1-2{y\over x}\right)\left[\beta-\beta_0\right] \right. \nn\\ 
&& \left. \qquad + 
{2 m_\pi^2 \over x} \left[\log\left({1+\beta_0 \over 1-\beta_0}
\right)-\log\left({1+\beta \over 1-\beta }\right) \right]
+ 2{y\over x} \right\rbrace~,
\eea
\be
\mbox{where} \qquad
\beta_0=\sqrt{1-{4m_\pi^2\over x }}\qquad\mbox{and}\qquad
\beta=\sqrt{1-{4m_\pi^2\over ( x-2y )}}~.
\ee

\end{document}